\documentclass[conference]{IEEEtran}
\ifCLASSINFOpdf
\else
\fi
\usepackage[utf8]{inputenc}
\usepackage{amsmath}
\usepackage{amsfonts}
\usepackage{amssymb}
\usepackage{graphicx}
\usepackage{subfigure}
\usepackage{float}
\usepackage{mathrsfs}
\usepackage{algorithm}
\usepackage[noend]{algpseudocode}
\usepackage{graphicx}
\usepackage{xcolor}
\usepackage{colortbl}
\usepackage{pbox}
\usepackage{color}
\usepackage{lipsum}
\usepackage[T1]{fontenc}
\usepackage[utf8]{inputenc}
\usepackage{url} 
\IEEEoverridecommandlockouts

\def\href#1#2{#1 #2}

\ifCLASSINFOpdf

\else

\fi

\begin{document}

\title{\huge Spatio-Temporal Motifs for Optimized \\ Vehicle-to-Vehicle (V2V) Communications \vspace{-0.5cm}}

\author{\IEEEauthorblockN{Tengchan Zeng\IEEEauthorrefmark{1}, Omid Semiari\IEEEauthorrefmark{2}, and Walid Saad\IEEEauthorrefmark{1}}
\IEEEauthorblockA{\IEEEauthorrefmark{1}Wireless@VT, Bradley Department of Electrical and Computer Engineering, Virginia Tech, Blacksburg, VA, USA}
Emails:\{tengchan, walids\}@vt.edu

\IEEEauthorblockA{\IEEEauthorrefmark{2}Department of Electrical Engineering, Georgia Southern University, Statesboro, GA, USA} Email: osemiari@georgiasouthern.edu \vspace{-0.6cm}
\thanks{This research was supported by the U.S. National Science Foundation under Grants CNS-1513697 and IIS-1633363.}}
\maketitle
\begin{abstract}	
	
Caching popular contents in vehicle-to-vehicle (V2V) communication networks is expected to play an important role in road traffic management, the realization of intelligent transportation systems (ITSs), and the delivery of multimedia content across vehicles. However, for effective caching, the network must dynamically choose the optimal set of cars that will cache popular content and disseminate it in the entire network. However, most of the existing prior art on V2V caching is restricted to cache placement that is solely based on location and user demands and does not account for the large-scale spatio-temporal variations in
V2V communication networks. In contrast, in this paper, a novel spatio-temporal caching strategy is proposed based on the notion of temporal graph motifs that can capture spatio-temporal communication patterns in V2V networks. It is shown that, by identifying such V2V motifs, the network can find sub-optimal content placement strategies for effective content dissemination across a vehicular network. Simulation results using real traces from the city of Cologne show that the proposed approach can increase the average data rate by $45\%$ for different network scenarios.\vspace{-0.21cm}

\end{abstract}

\IEEEpeerreviewmaketitle

\section{Introduction}
Vehicle-to-vehicle (V2V) communication is seen as one of enabling technologies for intelligent transportation systems and a key enabler for many smart road and traffic management systems \cite{Geo}, as it allows critical information dissemination. Moreover, spurred by the availability of in-vehicle infotainment (IVI) systems disseminating entertainment and information content to passengers \cite{YWu}, there is a strong need for a high-speed and stable delivery of large multimedia files, such as videos, photos and songs, to the various cars within a vehicular network. For effective dissemination of such diverse content across vehicular networks, there is a need for effective content placement strategies to maximize the throughput of the system \cite{DR}. Moreover, to reap the benefits of V2V content dissemination, we must address different challenges including optimal cache placement and resource allocation \cite{MM}.\setlength{\parskip}{0pt}

Cache placement in V2V networks has recently attracted significant attention such as in \cite{Haowu}-\cite{Shosny2}. 
In such works, the popular contents are offloaded to the storage of a number of well-chosen cars and devices at off-peak hours in order to serve requests during peak traffic hours. In these scenarios, cars and devices that do not have the cached content will not have to download the content from wireless base stations (BSs). Instead, they can request the content directly from other cars having cached data which can eventually lead to a reduction in the traffic load at the BSs. Meanwhile, caching using local storage can reduce latency due to a shorter communication distance. However, the benefits of caching are highly dependent on the set of cars and devices chosen for caching the popular contents \cite{Smitha}.

In \cite{Haowu}, the authors use vehicle mobility data for content dissemination and combine the idea of opportunistic forwarding, trajectory based forwarding and geographical forwarding to develop a mobility-centric algorithm to place content in vehicular networks.
Meanwhile, the work in \cite{Guan} applies available users mobility patterns to develop a polynomial-time solution to maximize the saved cost by caching contents in local storage.
Moreover, in \cite{Smitha}, the authors use location information and subscription-based information to divide vehicles into different groups and, then, design a spatio-temporal multicast routing protocol to construct an optimized dissemination mesh network. The work in \cite{Shosny2} presents an optimal caching policy by using both mobility information and users' demands and proposes a greedy caching algorithm with polynomial order complexity to obtain bounds of the caching policy whose complexity grows exponentially with the number of users.
However, existing works, including \cite{Haowu}-\cite{Shosny2}, do not take into account the temporal dynamics of V2V communication networks, such as the frequency of occurrence of different V2V links, which can be a key metric for content placement. For example, using only location information to select the set of cars that will cache the content can lead to choosing cars which are unable to communicate with each other or with other cars, effectively limiting the benefits of caching.\setlength{\parskip}{0pt}

The main contribution of this paper is a novel framework for spatio-temporal caching in vehicular networks that is cognizant of intrinsic spatial and temporal patterns in V2V communications. In particular,
using the tools of \emph{temporal network analysis} \cite{UAlon}, we identify temporal motifs in the V2V network as key communication patterns observed among vehicles that appear more frequently compared with what is expected in a baseline, randomized reference system. After identifying the spatial-temporal motifs, the proposed approach then finds the best candidate cars for content placement. To our best knowledge, this is \emph{the first work that exploits spatial-temporal motifs for optimizing content dissemination in vehicular networks}. Simulation results using real traces from the city of Cologne, Germany, show that the proposed motif-based approach yields significant performance gains in terms of the average data rate, compared to a conventional location-based scheme. 

The rest of the paper is organized as follows. Section \uppercase\expandafter{\romannumeral2} presents the system model and problem formulation. In Section  \uppercase\expandafter{\romannumeral3}, we present the proposed motif-based approach. Section \uppercase\expandafter{\romannumeral4} provides the simulation results. Conclusions are drawn in Section \uppercase\expandafter{\romannumeral5}.\vspace{-0.1cm}

\section{System Model and Problem Formulation}\vspace{-0.1cm}
\begin{figure}[!t]
\centering
\includegraphics[width=3.5in,height=1.8in]{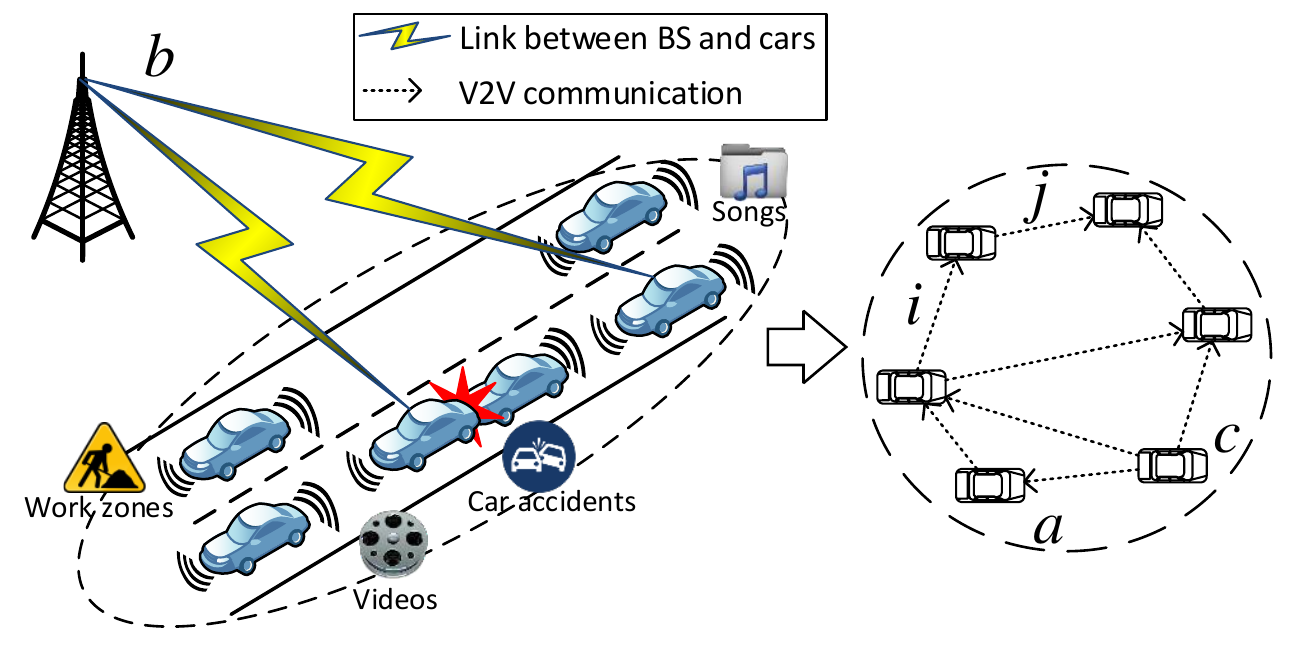}\vspace{-0.4cm}
\DeclareGraphicsExtensions.
\caption{V2V communication networks and its macroscopic communication graph. \vspace{-0.6cm}}\vspace{-0.4cm}
\label{systemmodel}
\end{figure}

Consider a vehicular network in an urban environment, composed of a set $\mathcal{N}$ of $N$ vehicles. In particular, as shown
in Fig.\ref{systemmodel}, we consider that a nearby BS $b$ provides wireless coverage to the cars in $\mathcal{N}$. The BS frequently uses beacon signals to keep track of the vehicles' positions by using the received signal-strength measurements in \cite{Martin}, and collects the V2V communication data when exchanging information with the communication facility within the vehicles as proposed in \cite{Hubaux}.

In our system, the BS seeks to seed multimedia files and traffic data in the storage unit of a set $\mathcal{C}\!\!\subseteq\!\!\mathcal{N}$ of $C$ cars to provide the passengers in neighboring cars with streaming services or to disseminate delay-sensitive information, such as upcoming road incidents to help drivers decide on their routes. Using this mechanism, the seeded cars, refereed to hereinafter as \emph{serving cars}, can disseminate the cached content to nearby cars via V2V links. Consequently, the BS can reduce its traffic load, as it no longer needs to transmit the same content to multiple cars. Fig. \ref{systemmodel} shows a case in which the BS caches content at car $c$. Here, an arbitrary car $a \!\!\in \!\!\mathcal{N}\!\setminus\!\! \{c\} $ sends a request for content $m$ to the BS $b$.
If the content is already cached at car $c$, the BS would inform the requesting car $a$ to use V2V communication to obtain the content from $c$.
Otherwise, the BS has to directly transmit the requested content to $a$, which increases its load. Hereinafter, we refer to cars that do not have the cached content as \emph{non-serving cars}. To increase the spectrum efficiency, we assume that V2V links reuse the spectral resources of the cellular network as in \cite{Yu}.\vspace{-0.1cm}

\subsection{Problem formulation}
The signal-to-interference-plus-noise ratio (SINR) of a V2V communication link between a serving car $c$ and a non-serving car $a$ will be:
\begin{align}\label{1}
\gamma_{c,a}=\frac{P_{c,a}g_{c,a}}{I_{b,a}+I_{a}^{\prime}+\sigma^{2}},
\end{align}
where $P_{c,a}$ is the transmission power from $c$ to $a$, and $\sigma^{2}$ is the variance of the Gaussian noise at the receiver. In addition, $g_{c,a}$ represents the channel gain between $c$ and $a$ and can expressed as $g_{c,a} \!= \! \eta d_{c,a}^{-\alpha}$, where $\eta$ is the fading gain, $d_{c,a}$ is the distance, and $\alpha$ is the path loss exponent.
We consider that the cellular links and the V2V links will experience a Rayleigh fading as done in \cite{Fang}. 
$I_{b,a}$ and $I_{a}^{\prime}$ capture, respectively, the interference generated by the links between the BS and other vehicles and by other V2V communication links. 
These interference terms are given by,
\begin{align}\label{2}
I_{b,a}\!\!= \!\!\sum_{a^{\prime}\in \mathcal{N}\setminus a}\!\!(L_{b,a^{\prime}}) P_{b,a^{\prime}}g_{b,a}, 
I_{a}^{\prime}\!\!=\!\!\sum_{c_{1}, a_{1}\in \mathcal{N}\setminus a}\!\! (L_{c_{1},a_{1}})  P_{c_{1},a_{1}}g_{c_{1},a}\,
\end{align}
where $L_{i,j}$ is a binary variable that captures the feasibility of  communication link between $i$ and $j$. In fact, $L_{i,j}=1$ if the SINR at the receiving car $j$ from the transmitting car $i$ exceeds a target threshold for V2V communication,
otherwise, $L_{i,j}=0$.

Accordingly, the achievable data rate for the V2V link between the serving car $c$ and non-serving car $a$ is 
\begin{align}\label{5}
R_{c,a}=\omega\log_{2}(1+\gamma_{c,a}),
\end{align}
where $\omega$ is the bandwidth. The corresponding SINR $\gamma_{b,a}$ and the achievable rate $R_{b,a}$ between the BS and non-serving car $a$, are:
\begin{align}\label{51}
\gamma_{b,a}= \frac{P_{b,a}g_{b,a}}{I_{a}^{\prime}+\sigma^{2}}, 
\vspace{1in}R_{b,a}=\omega\log_{2}(1+\gamma_{b,a}).
\end{align}

We consider a set $\mathcal{M}$ of $M$ popular contents with the same size, sorted from high to low popularity. Due to the limited storage, the serving nodes will choose to cache a limited number of contents. Therefore, to increase the possibility of meeting the requirements from non-serving nodes, the serving cars will cache $F\!\! \leq \!\!M$ most popular files, under the capacity constraints. 
Therefore, when a non-serving node requests one file out of the $F$ files, the node can acquire it via a V2V link as long as there exists one serving node within its communication range.  
To leverage V2V communications for disseminating cached content, the network must determine which vehicles act as serving nodes so as to maximize the average data rate achieved by non-serving cars. To this end, we define the binary variable $x_i$ for each car $i$, where $x_{i}\!\!=\!\!1$ if car $i$ is selected as serving node, otherwise $x_{i}\!\!=\!\!0$, and set $\mathcal{C}$ contains $C\!\!=\!\!\sum_{i}^{N}\!\!x_{i}$ serving nodes. Therefore, we can formulate the problem as follows:
\begin{align} \label{11}
 &\max_{\mathcal{C} \subseteq \mathcal{N}}\!\!\frac{1}{(N\!\!-\!\! C)}\! \!\sum_{a \in \mathcal{N}\setminus \mathcal{C}}\! \! \left(\sum_{m=1}^{F}\text{Pr}_{a}(m) \sum_{c\in \mathcal{C}}\beta_{c,a}R_{c,a}\!+\!\! \!\! \!\sum_{m=F+1}^{M}\! \!\! \!\text{Pr}_{a}(m)R_{b,a}\right)      \\ \vspace{-0.4cm}
&\text{s.t.}\hspace{0.1in}
P_{c,a}\leq P_{\text{max}}, \label{size} \\
&\hspace{0.3in}
\bar{\gamma}\leq\gamma_{c,a}, \label{1111}
\hspace{0.4in}
\end{align}
where $\text{Pr}_{a}(m)$ is the probability mass function (pmf) of the request for file $m$ by car $a$.  This distribution can be modeled by the Zipf distribution with
pmf $\text{Pr}_{a}(m)\!\! =\!\! \frac{1}{m^{\theta_{r}}}/\sum_{x=1}^{M}\frac{1}{x^{\theta_{r}}}$, where $\theta_{r}$ is the Zipf exponent that determines the skewness of the
distribution \cite{Malak}. The indicator variable $\beta_{c,a}$ is such that $\beta_{c,a}\!\!=\!\!1$ if the serving car $c$ is the nearest car to the requester $a$ in the serving set $\mathcal{C}$; $\beta_{c,a}\!\! =\!\!0$, otherwise. $\beta_{c,a}$ ensures that a non-serving car will always choose the closest serving car $c$ with the requested file.
Constraint (\ref{size}) guarantees that the transmission power of vehicles will not surpass the maximum power level $P_{\text{max}}$, and 
constraint (\ref{1111}) ensures that SINR of the V2V links is above a threshold, $\bar{\gamma}$. \vspace{-0.1cm}

\subsection{V2V macroscopic communication graphs}

As long as we find the optimal set of serving vehicles, we can assign non-serving nodes to serving nodes according to the spatial distance, and solve the problem given by (\ref{11})-(\ref{1111}). However, finding the optimal set $\mathcal{C}$
is a 0-1 integer programming where determining whether each individual node in the set $\mathcal{N}$ should be considered as either a serving or a non-serving node. In fact, the problem is one of Karp’s 21 NP-complete problems \cite{Karp}, which is hard to solve directly.


Alternatively, we can find a sub-optimal solution to the optimization problem by using the information within the vehicular network. In particular, in addition to the the position and demand of each vehicle, we can also leverage the temporal domain information. 
This is because V2V networks are naturally dynamic and exhibit the temporal features. For example, the number of communication links between two arbitrary cars may vary with time. Such temporal information on the frequency of communication is valuable to determine which cars are more likely to better disseminate the content. Therefore, time domain information is also important to choose the optimal set $\mathcal{C}$ for cache placement and solve the problem (\ref{11})-(\ref{1111}).

To capture the dynamics in the time domain, we propose to use collected V2V communications data and model the system as a directed \emph{temporal graph} $G \!= \!(\mathcal{N}, \mathcal{E})$, whose vertices are the cars in $\mathcal{N}$ and whose temporal edges, in set $\mathcal{E}$, denote the time-stamped communication events among vehicles. We represent a wireless communication link between two different cars
as a \emph{3-tuple} edge labeled as $i$, $e_{i}\!=\!<\!c_{i},a_{i},t_{i}\!>\!$, where $c_{i}$ and $a_{i}$ denote the serving car and the non-serving car, respectively. $t_{i}$ is the time of initiating the transmission from $c_{i}$ to $a_{i}$. Here, we assume that, at any time, a car cannot communicate with more than one car simultaneously. In this case, there are no two edges with the common vehicle element initiated at the same time.  

To obtain the set of cars that are more active and more likely to participate in the V2V communication in a period of time, we introduce the time constraint $T$ and devide the graph $G \!= \!(\mathcal{N}, \mathcal{E})$ into multiple \emph{macroscopic communication graphs}, where we can change the value of $T$ to filter outdated V2V links. That is, for any edge $e_{i}$ in these macroscopic graphs, there always exists at least one other edge $e_{j}, i\neq j$, in the same graph, that meets the following requirements:
\begin{itemize}
  \item Two edges share at least one node, $(c_{i},a_{i}) \bigcap (c_{j},a_{j}) \neq \emptyset$;
  \item If a wireless connection $i$ occurs before another connection $j$, $0 \leq t_{j}-t_{i} \leq T$;
  \item If a wireless connection $j$ occurs before another connection $i$, $0 \leq t_{i}-t_{j} \leq T$.
\end{itemize}

Given these definitions, using the graph-theoretic framework of \emph{temporal networks analysis}, next, we present a novel graph-theoretic sub-optimal solution to the original problem in (\ref{11})-(\ref{1111}).\vspace{-0.1cm}

\section{Proposed Strategy Based on Spatio-Temporal Motifs}\vspace{-0.1cm}
\begin{algorithm}
\caption{ Microscopic Subgraphs Searching Algorithm}\label{euclid}
\textbf{Input:} A V2V macroscopic communication graph $G_{1}=(\mathcal{N}_{1},\mathcal{E}_{1})$, and target subgraph size $k$. \\
\textbf{Output:} The set of microscopic subgraph $\mathcal{V}_{1}$ with size $k$.
\begin{algorithmic}[1]\footnotesize
\State  $\mathcal{V}_{1}=\varnothing$
\State \textbf{for} {arbitrary edge $e_{i} \in \mathcal{E}$ } \textbf{do}
\State \hspace{10pt} $\mathcal{V}_{1}\leftarrow \{e_{i}\}$
\State \hspace{10pt} \textbf{for} edge $e_{j}\in \mathcal{E}-e_{i}$  \textbf{do}
\State \hspace{20pt} \textbf{if} $\{c_{i},a_{i}\}\cap\{c_{j},a_{j}\}\neq \varnothing$ \textbf{and} $j>i$
\State \hspace{30pt} $\mathcal{V}_{2}\leftarrow \{e_{j}\}$
\State \hspace{20pt} \textbf{end}
\State \hspace{10pt} \textbf{end}
\State \hspace{10pt} \textbf{call} \textproc{EdgeExtension}($\mathcal{V}_{1}, \mathcal{V}_{2}, k$)
\State \textbf{return} $\mathcal{V}_{1}$
\State \textbf{function} \textproc{EdgeExtension}($\mathcal{E}_{\text{sub}}$, $\mathcal{V}_{3}$, $k$)
\State  \textbf{If} $|\mathcal{E}_{\text{sub}}|= k$ \textbf{then output} $G[\mathcal{E}_{\text{sub}}]$ and \textbf{return}
\State \hspace{0pt} \textbf{while} $\mathcal{V}_{3}\neq \emptyset$ \textbf{do}
\State \hspace{10pt} \textbf{for} arbitrary edge $e_{x} \in \mathcal{V}_{3}$  \textbf{do}
\State \hspace{20pt} Remove $e_{x}$ from $\mathcal{V}_{3}$
\State \hspace{20pt} \textbf{for} edge $e_{y}\in \mathcal{V}_{3}-e_{x}$  \textbf{do}
\State \hspace{30pt} \textbf{if} $\{c_{x},a_{x}\}\cap\{c_{y},a_{y}\}\neq \varnothing$ \textbf{and} $y>x$
\State \hspace{40pt} $\mathcal{V}_{4}\leftarrow \{e_{y}\}$
\State \hspace{30pt} \textbf{end}
\State \hspace{20pt}\ \textbf{end}
\State \hspace{20pt} $\mathcal{V}_{3}^{\prime}=\mathcal{V}_{3} \cup \mathcal{V}_{4}$
\State \hspace{20pt} \textbf{call} \textproc{EdgeExtension}($\mathcal{E}_{\text{sub}}\cup \{e_{x}\}$, $\mathcal{V}^{\prime}_{3}$, $k$)
\State \hspace{10pt}\ \textbf{end}
\State  \textbf{return}
\newline
\vspace{-0.4cm}
\end{algorithmic}
\end{algorithm}
\setlength{\parskip}{0pt}

To find a sub-optimal solution to (\ref{11})-(\ref{1111}), we need to develop an algorithm to detect motifs in V2V communication macroscopic graphs. To this end, we follow three steps. The first step is searching \emph{microscopic subgraphs} defined as basic communication units with typical sizes (number of edges) in macroscopic communication graphs. The second step is collecting microscopic subgraphs with similar isomorphic structure.
The third step is determining the frequently occurring motifs after comparing each subgraph's frequency $f$ of occurrence with its counterpart $f^{\prime}$ in a baseline randomized V2V communication network. In particular, we use the notion of \emph{Z-score}, expressed as \vspace{-0.1cm}
\begin{align}\label{zscore}
Z=\frac{f-f^{\prime}}{\sigma_{m}},
\end{align}
where $\sigma_{m}$ captures the standard deviation of the corresponding subgraph in the reference system. If $Z\!\! >\!\! Z_{th}$, where $Z_{th}$ is a given threshold, we can classify the subgraph as a motif \cite{Temporal}. Finally, 
given the detected motifs, our spatio-temporal caching strategy can select $\mathcal{C}$, $\mathcal{C}\!\! \subseteq \!\!\mathcal{N}$, to solve the optimization problem in (\ref{11})-(\ref{1111}). \vspace{-0.4cm}
\begin{figure*}[!t]
\centering
\includegraphics[width=5.5in]{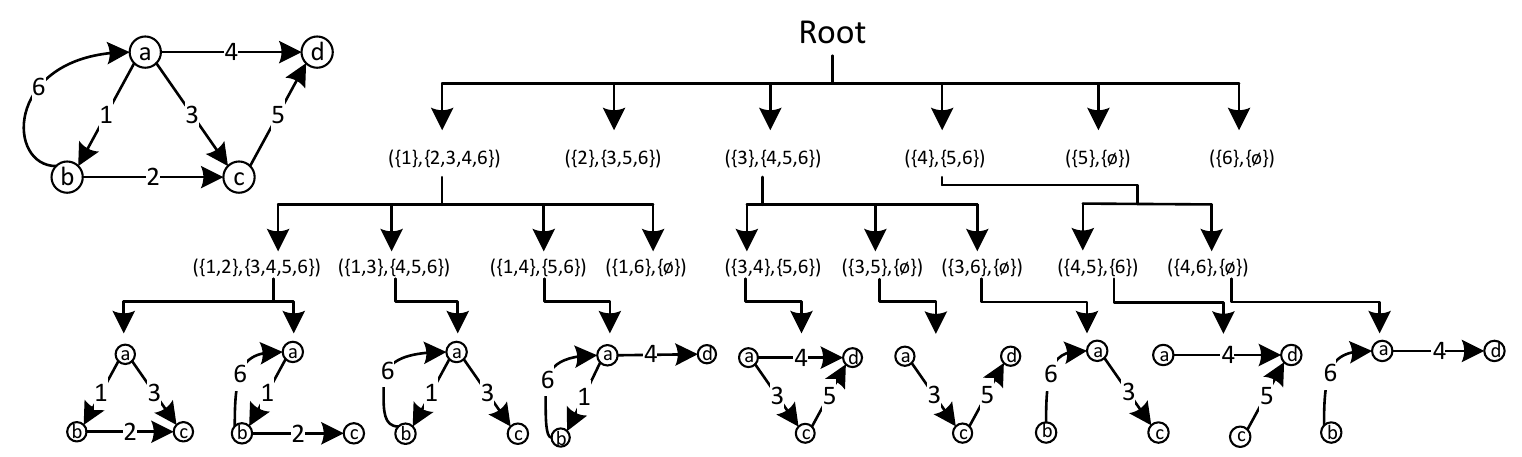}\vspace{-0.3cm}
\DeclareGraphicsExtensions.
\caption{Considering a graph with $4$ nodes and $6$ edges in the corner, we can obtain all microscopic subgraphs with size as $3$ after following Algorithm 1.}
\vspace{-0.7cm}
\label{systemmodel111a}
\end{figure*}
\vspace{-0.2cm}

\subsection{Searching for V2V microscopic subgraphs}
To effectively detect existing motifs in V2V communication networks, we first decompose V2V macroscopic communication graphs into microscopic subgraphs with the same size.

The algorithm used for finding V2V microscopic subgraphs with target size $k$ in a macroscopic graph $G_{1} \!= \!(\mathcal{N}_{1}, \mathcal{E}_{1})$ is shown in Algorithm \ref{euclid}. Given a set of labeled cars and labeled connection edges, there are two ways to obtain microscopic subgraphs from the macroscopic graph. One approach is to consider the nodes connected by edges, and another approach is to consider the set of edges in which an arbitrary edge can find another edge sharing the common node. In contrast to the work in \cite{SW}, the proposed algorithm is based on the second approach. As shown in Fig. \ref{systemmodel111a}, from the macroscopic graph, we can obtain a pair of edge sets, $\mathcal{V}_{1}$ and $\mathcal{V}_{2}$, where the first set contains only one edge $e_{i}$, labeled as $i$, and the second set has $e_{i}$'s all neighboring edges with greater labels.
The mechanism will first add an arbitrary edge $e_{j}$, labeled as $j$, from the second set to the first set and update the first set. By calling \textproc{EdgeExtension($\mathcal{E}_{\text{sub}}$, $\mathcal{V}_{3}$, $k$)}, the second set could be extended by first merging the set of edges, which are neighbor to the newly added edges $e_{j}$ and with greater labels in the macroscopic graph, and then removing edge $e_{j}$ and other edges with smaller labels compared with $j$.
Next, we repeat the aforementioned steps for the first set with the added edge and the new second set until we obtain the microscopic subgraphs meeting the size requirement, i.e., $k$.

\textbf{Example:} When the two sets of edges from the macroscopic graph are \{1\} and \{2,3,4,6\}, as shown in Fig. \ref{systemmodel111a}, the algorithm first adds edge 2 to the first set \{1\} and obtains the first updated set \{1,2\}. Then, it updates the second set as \{3,4,5,6\} by first merging the neighboring set with greater labels, i.e., \{3,5,6\} with \{2,3,4,6\} and then deleting the edge 2. Repeatedly, the algorithm follows the same processes for the updated set \{1,2\} and the edge set \{3,4,5,6\}, and we can obtain the microscopic subgraphs \{1,2,3\} and \{1,2,6\}, if $k=3$.

As shown in Fig. \ref{systemmodel111a}, by using this algorithm for any one edge set and the set of its surrounding edges with greater labels, we can finally collect all V2V microscopic subgraphs with the required size in a macroscopic graph. To obtain the whole set of microscopic subgraphs existing in the V2V network, we can use this algorithm for different macroscopic graphs. Although the complexity of the algorithm will increase with the increment of the vehicles, we can apply the algorithm in scenarios with capacity limitations, like intersections,  parking lots, and parts of the highway, or a subset of the network, thus reducing the complexity and the processing time. \vspace{-0.1cm}

\subsection{Classifying V2V microscopic subgraphs}

\begin{figure}[!t]
\centering
\includegraphics[width=2.8in,height=1.8in]{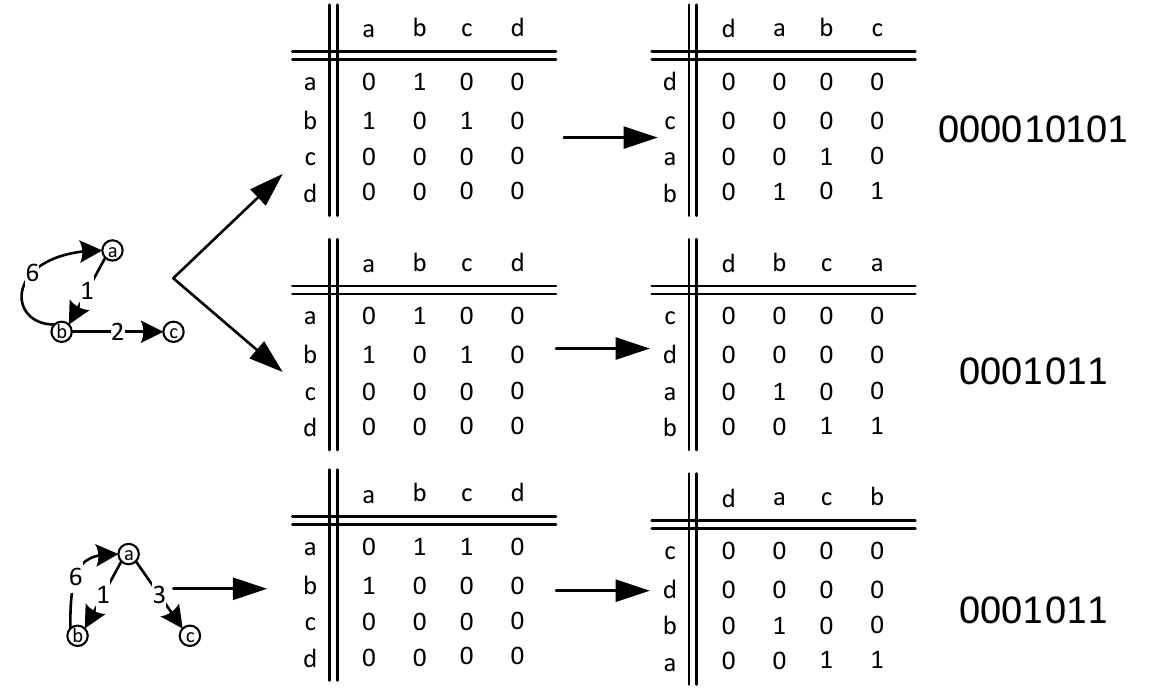}\vspace{-0.4cm}
\DeclareGraphicsExtensions.
\caption{Examples on the canonical labeling of a microscopic subgraph and other microscopic subgraphs sharing the same canonical labeling.}
\label{systemmodel2}\vspace{-0.6cm}
\end{figure}

To sort V2V microscopic subgraphs with the same isomorphic structure, i.e., to find microscopic subgraphs which contain the same number of cars connected in the same way, we exploit the notion of \emph{canonical labeling}. By permuting the elements in the adjacency matrices obtained from a microscopic subgraph, we could construct many lists of integers. By viewing each list of integers as a string of 1s and 0s, we sort them based on lexicographic ordering to obtain a \emph{canonical labeling} defined as the string with the minimum value. Due to the fact that the canonical labeling of two subgraphs will be identical as long as they have the similar isomorphic structure \cite{R.C}, the problem of
determining isomorphic structures among subgraphs is equivalent to deciding whether given microscopic subgraphs
have the same canonical labeling or not.

For example, as shown in Fig. \ref{systemmodel2}, by concatenating rows or columns one after the other in the permuted adjacency matrices of the first microscopic extracted from Fig. \ref{systemmodel111a}, we can find two strings of 1s and 0s.
According to lexicographic ordering, we notice the string corresponding to the second matrix is smaller than its counterpart in the first matrix, i.e., ``000010101''<``0001011'', and ``0001011'' can be chosen as the canonical labeling for the first microscopic subgraph compared with other permutations. Similarly, we can observe the second microscopic subgraph shares the same canonical labeling with the first subgraph, and, thus, these two subgraphs have the same isomorphic structure.

Accordingly, we are capable of completing structure classification for all the microscopic subgraphs in the V2V communication network, and then, we can calculate the occurrence frequency $f$ of microscopic subgraphs having a similar isomorphic structure.
Moreover, we also repeat the same steps for a randomized V2V communication network so as to obtain the mean frequency $f^{\prime}$ and the standard deviation $\sigma_{m}$ of the corresponding subgraph. We can use (\ref{zscore}) to determine whether the subgraph is motif or not.  \vspace{-0.25cm}

\subsection{Proposed spatio-temporal caching strategy}
\begin{figure}[!t]
\centering
\includegraphics[width=3in,height=2.6in]{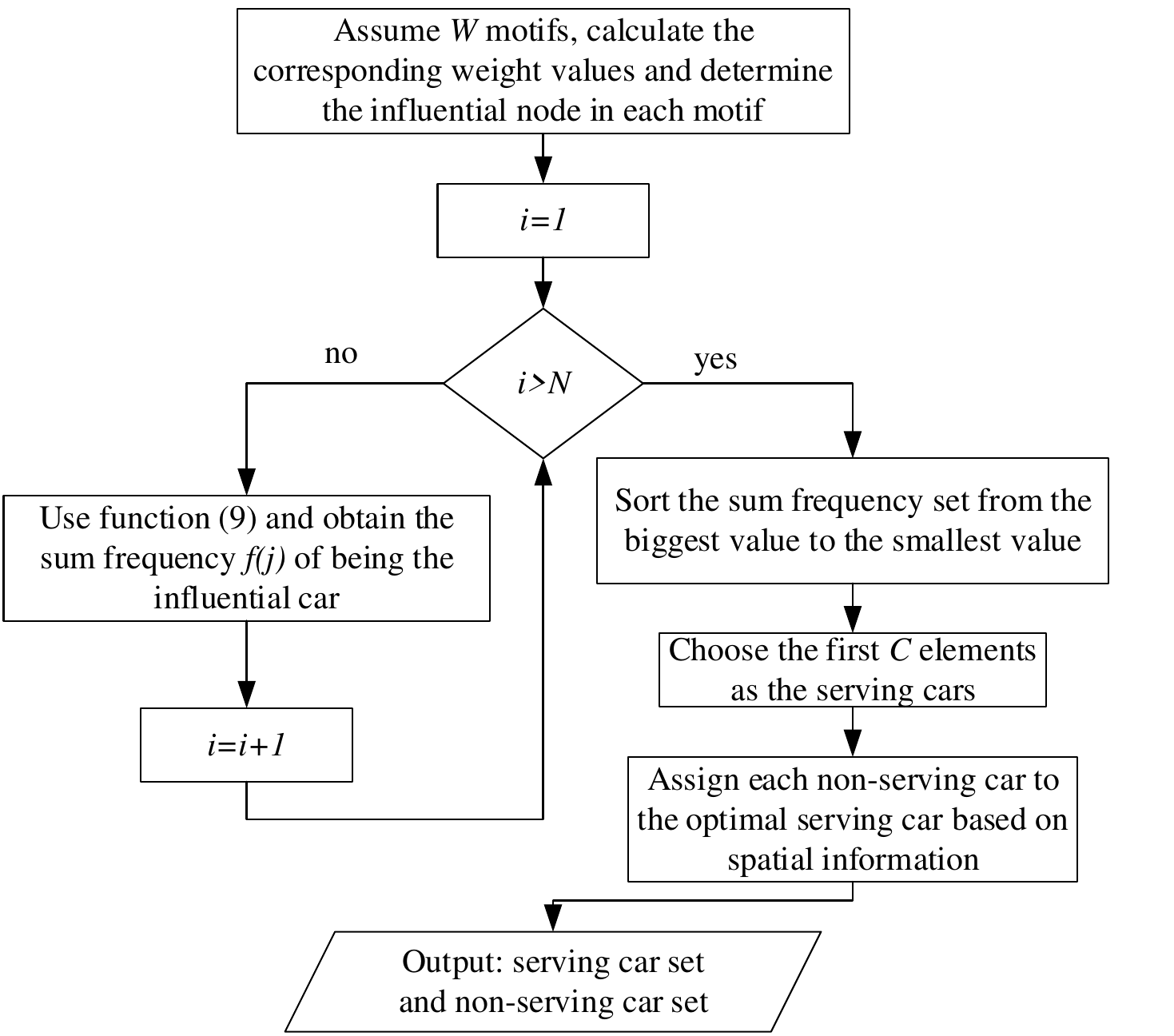}
\DeclareGraphicsExtensions.\vspace{-0.2cm}
\caption{Flow chat of the proposed spatio-temporal caching strategy.}
\label{Pro2}\vspace{-0.5cm}
\end{figure}
To obtain the sub-optimal solution for the optimization problem (\ref{11})-(\ref{1111}), as shown in Fig. \ref{Pro2}, the BS will first select the serving cars in the set $\mathcal{C}$ based on the motifs detected from the V2V communication graph. In particular, we assume there are $W$ motifs and the Z-score of a motif $i$ is $Z_{i}$. Note that for any $i\in \{1,...,W\}$, $Z_{i}>Z_{th}$. By observing the motif structure, we can determine the outdegree of each car, which is defined as the number of outgoing edges emanating from the car. For example, for the first microscopic subgraph in Fig. \ref{systemmodel2}, the outdegrees of $a, b, c$ are 1, 2, and 0. Then, we choose the influential car as the node with the maximum outdegree in the corresponding motif, since the connections that originate from this car can reach more recipients compared with other cars, leading to a more effective content dissemination. Next, we can statistically acquire the frequency of car $j$'s being the influential car in the $i-$th motif as $f_{ij}, j \in \mathcal{N}$. Based on that, as shown in Fig. \ref{Pro2}, we can obtain the sum frequency of being the influential car in the motifs for car $j$ as,\vspace{-0.2cm}
\begin{equation}\label{mott}
\setlength{\abovedisplayskip}{3 pt}
\setlength{\abovedisplayskip}{3 pt}
f(j)=\sum_{i=1}^{W} w_{i}f_{ij},
\end{equation}
where $w_{i}$ captures the weight value of a motif $i$, expressed as $w_{i}=\frac{Z_{i}}{\sum_{j=0}^{W}Z_{j}}$.
After calculating the sum frequency of each car, we sort nodes from the car with the highest frequency to the one with the lowest frequency. Finally, the choice of best $C$ candidates to cache will be the first $C$ elements in the array.

After choosing the serving cars, the next step is to assign each receiving car to its optimal serving car. In particular, if the BS receives the request from one receiving car, the BS would inform the car the nearest serving car with required content based on the collected location information. To solve (\ref{11})-(\ref{1111}), we decompose it into two problems. One is determining the optimal set of serving cars, which is solved by exploiting the temporal motifs. The other is completing the best assignments between the serving cars and the non-serving cars using the spatial knowledge.\vspace{-0.1cm}

\section{Simulation results}\vspace{-0.1cm}

\begin{table}[!t]
    \large
	\begin{center}
		\caption{\small Simulation parameters.}
		\vspace{-0.2cm}
		\label{table_example}
		\resizebox{9cm}{!}{
			\begin{tabular}{|c|c|c|}
				\hline
				\textbf{Parameter} & \textbf{Meaning} & \textbf{Value} \\ \hline \hline	
				$ w_{l} $  & Width of each lane & $3.5$~m  \\ \hline
                $P_{b,n}$ &   Transmission power of base station     &  $20$~W     \\ \hline
				$P_{\text{max}}$ & Transmission power of V2V links  & $20$~dBm  \\ \hline
                $\alpha$  & Path loss exponent & $3$ \\ \hline
                $\bar{\gamma}$  & SINR threshold  & $10$~dB  \\ \hline
                $\sigma^{2}$ & Power of noise  & $-94$~dBm  \\ \hline
                $\omega$  &Bandwidth of the system  & $75$~MHz  \\ \hline
                $\theta_{r}$ &Zipf exponent & $2$ \\ \hline
                $M$  & Total number of files in the network   & $10$ \\ \hline
                $F$  & The maximum number of files the car can cache & $3$ \\ \hline
                $d_{b}$ &Approximate distance from BS to freeway & $10.0$~km \\ \hline
                $T$ & Time constraint & 100~s \\ \hline
			\end{tabular}}
		\end{center}\vspace{-0.7cm}
	\end{table}

For our simulations, we use the vehicular mobility dataset within the city of Cologne, Germany, which is collected by the TAPASCologne project \cite{TAPA}.
The dataset contains information about roads and vehicles as well as the trip information for each individual car in one day. 
In particular, we take into account a $5$ km-length freeway (Autobahn 4 in
Cologne) with three lanes in each direction and the nearest BS is colonius fernsehturm. We collect the location information of a network of $53$ vehicles that coexist on the freeway from the data.
All simulation parameters are summarized in Table. \ref{table_example}.

Since the Poisson distribution can be used to capture the number of events occurring within a fixed interval \cite{Poisson}, we assume that the number of wireless communication links between two arbitrary cars $i$ and $j$ that in proximity of one another, in a given period of time, follows a Poisson distribution with parameter $\lambda_{i,j}$. To better simulate real-time data, we assume that $\lambda_{i,j}$ is inversely proportional to the distance between $i$ and $j$. This is because a closer distance leads to a better communication environment, resulting in a higher probability to build  communication links. Then, we randomly assign a time stamp to each V2V communication links.
In a baseline randomized V2V communication network, the communication instances are randomly generated and given a time stamp.
Once the temporal graph data is generated, the motifs can be detected based on the approach in Section \uppercase\expandafter{\romannumeral3}. In our analysis, a structure is identified as a motif, when the Z-core of the structure is at least $2$.
For comparison, we use a location-based caching strategy. In this strategy, the basic principle is that the BS would always select a set of cars that can realize the least summation of distance between remaining cars and the corresponding closest cars the in the selected set. 

\begin{figure}[!t]
\centering
\includegraphics[width=2.0in,height=0.25in]{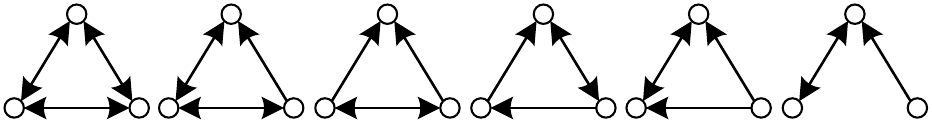}
\DeclareGraphicsExtensions.\vspace{-0.2cm}
\caption{Motifs found in the first scenario.}
\label{Average11}\vspace{-0.5cm}
\end{figure}
\begin{figure}[!t]
\centering
\subfigure[$31$ cars]{
\label{Motif1}
\includegraphics[width=1.33in,height=0.25in]{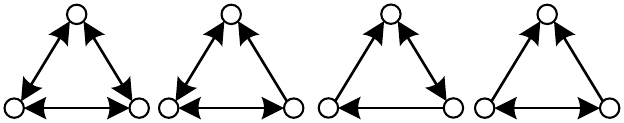}}\vspace{-0.1cm}
\subfigure[$33$, $35$, and $37$ cars]{
\label{Motif2}
\includegraphics[width=1.65in,height=0.25in]{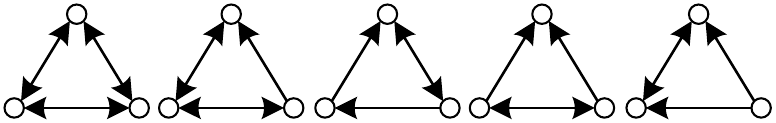}}\vspace{-0.0cm}
\subfigure[$39$ and $41$ cars]{
\label{Motif3}
\includegraphics[width=2.0in,height=0.25in]{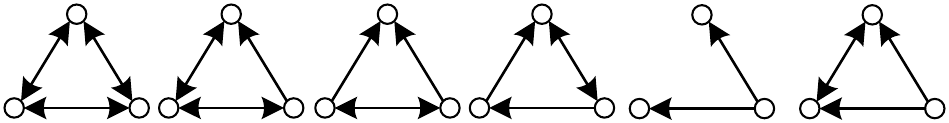}}\vspace{-0.1cm}
\subfigure[$43$, $45$, $47$, $49$, $51$ and $53$ cars]{
\label{Motif4}
\includegraphics[width=1.65in,height=0.25in]{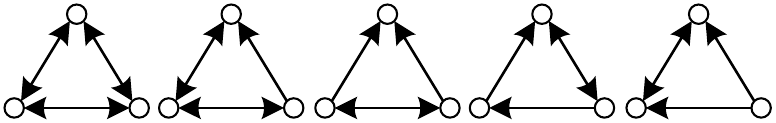}}\vspace{-0.1cm}
\caption{Motifs found in the second scenario with different number of cars.}
\label{fig:1}\vspace{-0.7cm}
\end{figure}

We consider two simulation scenarios. 
In the first scenario, we consider total number of $53$ cars (serving and non-serving). As we change the number of serving cars, the number of non-serving cars will be modified accordingly.
In the second scenario, we first randomly choose several car sets with different total number of cars. Then, based on the proposed strategy and the location-based strategy, we choose a fixed number of non-serving cars chosen out of the selected sets. In particular, we select twelve car sets having a total number of cars ranging from $31$ to $53$ with a step of $2$, and we fix the number of receiving cars to $30$.
According to the proposed method, we can detect different motifs in both scenarios, sorted from the highest to the lowest in terms of Z-score, from the real trace data and generated wireless communication data, as shown in Fig. \ref{Average11} and Fig. \ref{fig:1}.
Based on these structures, we can observe the outdegree of each node. Then, using the proposed algorithm in
Section \uppercase\expandafter{\romannumeral3}, we can determine the set of cars used for caching.

\begin{figure}[!t]
\centering
\includegraphics[width=2.89in]{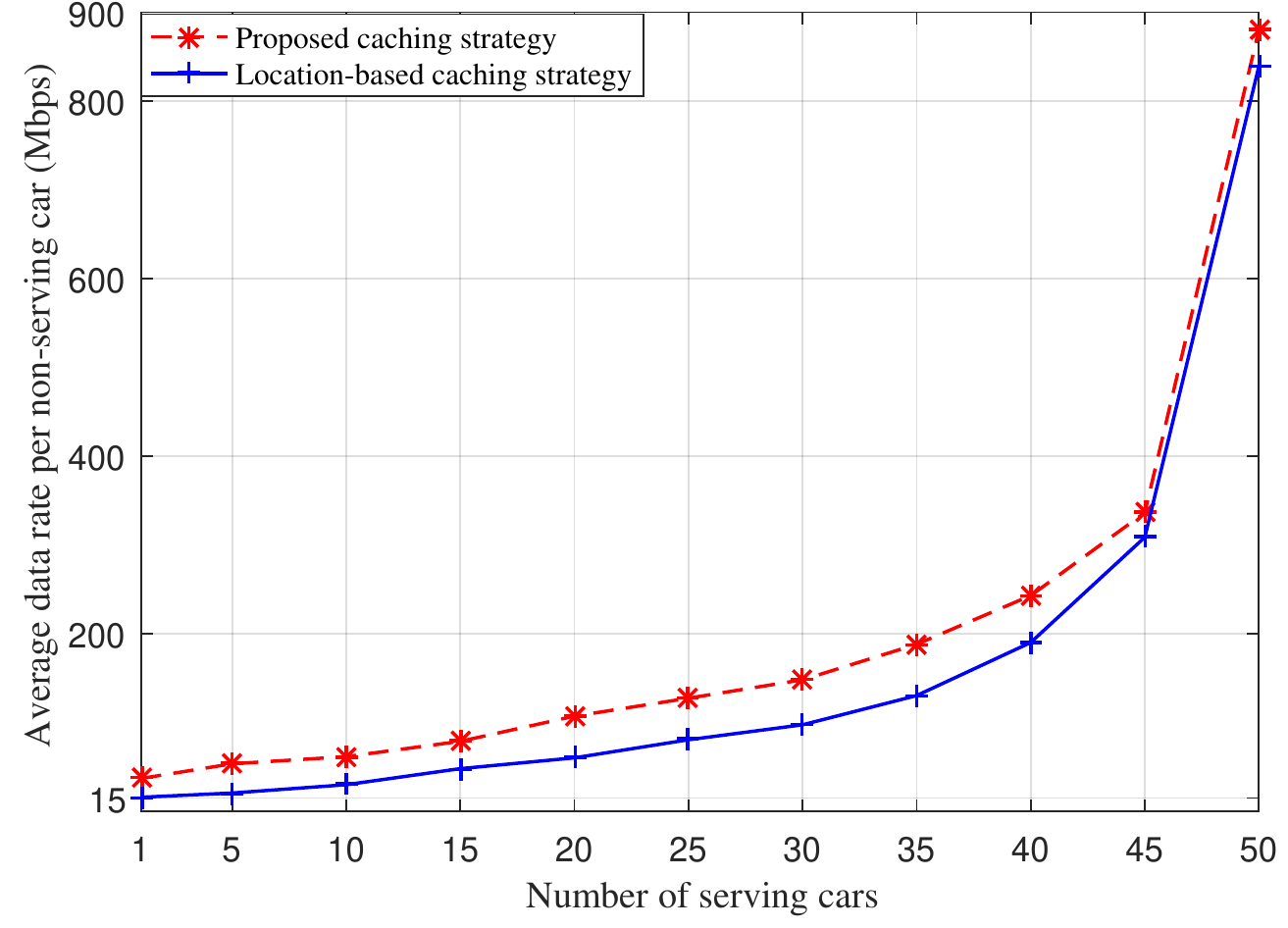}
\DeclareGraphicsExtensions.\vspace{-0.2cm}
\caption{Average data rate for non-serving cars where the number of cars is $53$.}
\label{Average}\vspace{-0.38cm}
\end{figure}

Fig. \ref{Average} shows the average transmission rate achieved by non-serving cars under the location-based caching strategy and the proposed approach, with the total number of cars fixed at $53$ (first scenario). From Fig. \ref{Average},
we observe that the spatio-temporal caching strategy yields a better performance compared with the location-based cache strategy in terms of average date rate per non-serving car. In particular, the performance advantage reaches up to $45\%$ when the number of serving car is $20$. Furthermore, as the number of serving cars increases, the number of non-serving cars will decrease. In particular, when the number of serving cars goes to $50$, there are only $3$ cars requesting for content, leading to a reduced interference and an increase in the average data rate, as also seen in Fig. \ref{Average}.

\begin{figure}[!t]
\centering
\includegraphics[width=2.89in]{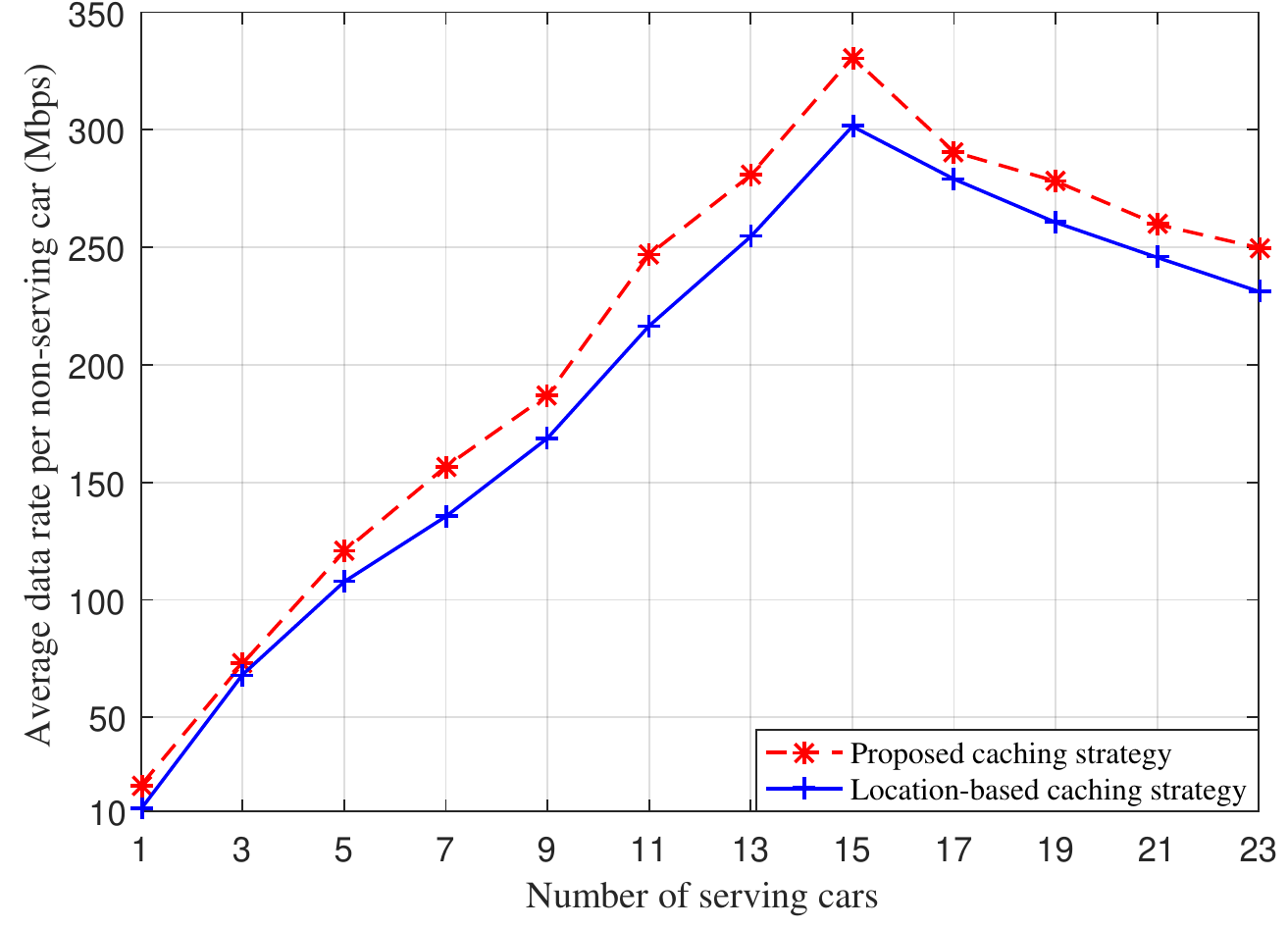}
\DeclareGraphicsExtensions.\vspace{-0.2cm}
\caption{Average data rate for non-serving cars where the number of non-serving cars is $30$.}
\label{Pro}\vspace{-0.38cm}
\end{figure}

\begin{figure}[!t]
\centering
\includegraphics[width=2.89in]{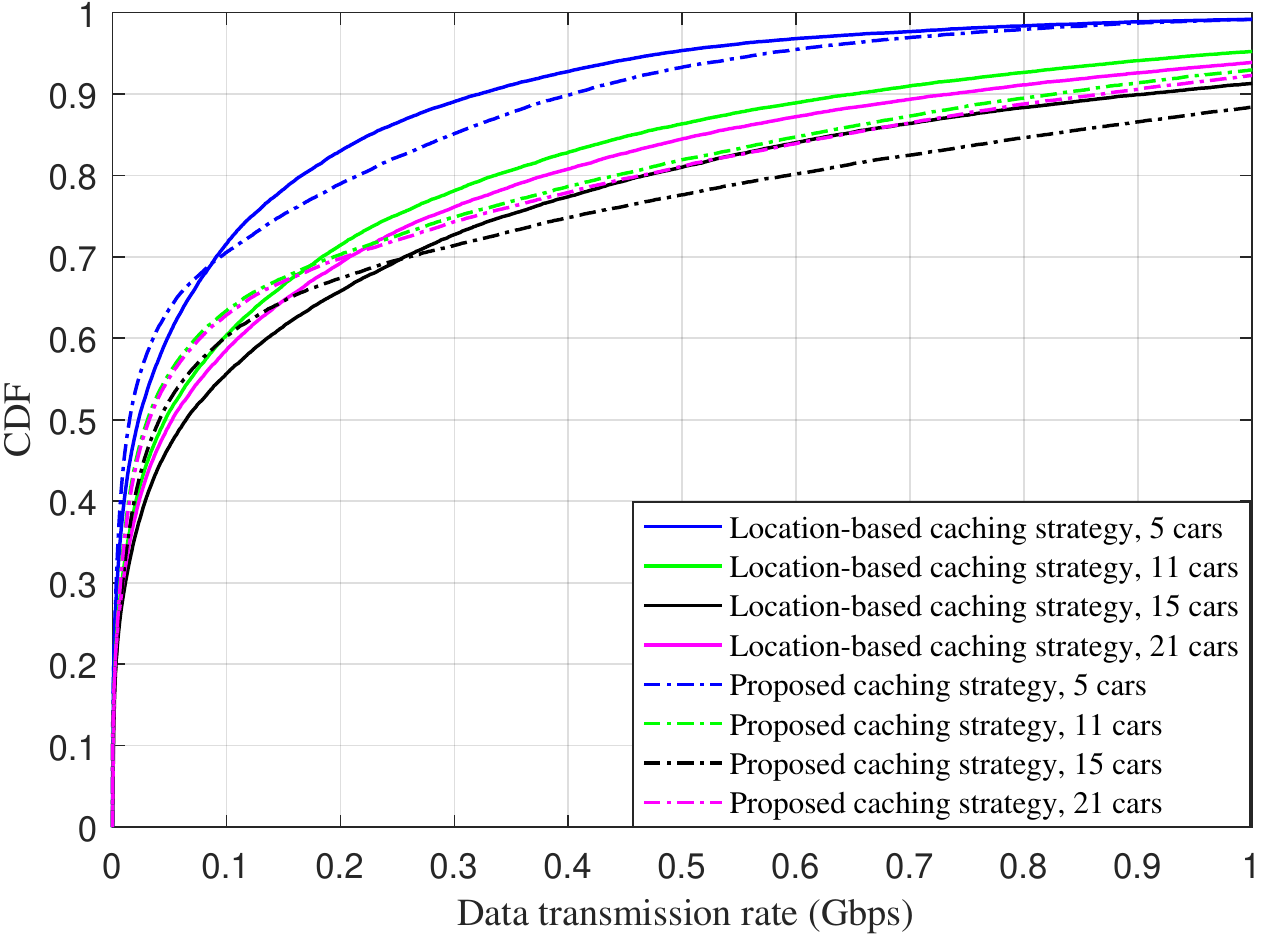}
\DeclareGraphicsExtensions.\vspace{-0.2cm}
\caption{CDFs for the data rate of non-serving cars where the number of serving cars are $5$, $11$, $15$, $21$.}
\label{Pro11}\vspace{-0.6cm}
\end{figure}

Fig. \ref{Pro} shows the average data rate for $30$ non-serving cars as we vary the number of serving cars. We can observe that the proposed strategy outperforms location-based caching strategy
up to $15\%$ when there are $11$ cars acting as serving nodes.
Further, the average data rate for both strategies will not increase all the time. This is due to the fact that an increase in the number of serving cars will raise the interference over the V2V links. When the impact of interference cannot be compensated by the gain from V2V communication, the average data rate for V2V links will decrease, as seen in Fig. \ref{Pro}. Moreover, Fig. \ref{Pro11} shows the cumulative distribution functions (CDFs) of the data rate of non-serving cars for $5$, $11$, $15$, and $21$ serving cars. Compared with the data rate resulting from location-based strategy, the non-serving cars are more likely to achieve a higher data rate when employing the proposed caching strategy. Fig. \ref{Pro11} also shows under both caching strategies, the probability of achieving a higher data rate for $15$ serving cars is greater than the counterparts for $5$, $11$, and $21$ serving cars. In particular, when the number of serving cars is $15$ and the probability is  $0.8$, the proposed caching strategy improves the data rate of about $20\%$ compared to the location-based strategy.\vspace{-0.3cm}


\section{Conclusions}
In this paper, we have proposed a novel spatio-temporal caching policy in vehicular networks. In contrast to traditional location-based caching strategies, we have leveraged temporal graph motifs, which represent subgraphs with high frequency of occurrence in the V2V communication graph, to determine car candidates to cache popular content. We have developed an approach to detect the motifs and, then, have used the results to determine the preferred set of cars for popular content placement. Simulation results using real car location traces have shown that the proposed spatio-temporal caching strategy can yield significant gains in terms of the average data rate per car for two practical scenarios.
\vspace{-0.5cm}
\def\baselinestretch{.88}

\end{document}